\documentclass{iopconfser}

\usepackage{epsfig,graphicx,rotating,color}

\newcommand{\ba}{\begin{eqnarray}}
\newcommand{\ea}{\end{eqnarray}}
\begin{document}

\title{Algebraic models of $\alpha$-cluster nuclei}

\author{Roelof Bijker$^{1}$}

\affil{$^1$Instituto de Ciencias Nucleares, Universidad Nacional Aut\'onoma de M\'exico, Ciudad de M\'exico, Mexico}

\email{bijker@nucleares.unam.mx}

\begin{abstract}
In this contribution I review the algebraic cluster model (ACM) for $\alpha$-conjugate nuclei with $A=4n$ and its extension to the cluster shell model (CSM) for $A=4n+x$ nuclei. 
The CSM describes the splitting of single-particles levels in a deformed field generated by the cluster of $\alpha$-particles. As an example, I discuss the energies, electromagnetic transitions and form factors of the neighboring nuclei $^{12}$C and $^{13}$C.
\end{abstract}

\section{Introduction}

The physics of light nuclei arises from the complex interplay of both collective cluster degrees of freedom and single-particle degrees of freedom, which presents a daunting challenge to nuclear structure models, {\it ab initio} calculations and nuclear effective field theory 
(see {\it e.g.} \cite{Oertzen,FreerFynbo,Freer} and references therein). In addition to the interest in the nuclear structure aspects light nuclei are also of crucial importance to nuclear astrophysics and nucleosynthesis due to the existence of near-threshold $\alpha$-cluster states \cite{Wiescher}. The renewed interest in the structure of $\alpha$-cluster nuclei was stimulated in part by the measurement of new rotational excitations of both the ground state \cite{Freer2007,Kirsebom,Marin} and the Hoyle state in $^{12}$C \cite{Itoh,Freer2012,Gai,Freer2011}. 

In this contribution I review the algebraic cluster model (ACM) for $\alpha$-cluster nuclei  with $A=4n$ and its extension to the cluster shell model (CSM) for $A=4n+x$ nuclei. 
The ACM provides a new theoretical approach for the study of clustering in $\alpha$-conjugate nuclei based on the discrete symmetry of the underlying geometric configuration of $\alpha$ particles. An interesting question is to what extent the $\alpha$-cluster structure survives under the addition of extra nucleons. The CSM describes the behavior of single-particles levels in a deformed field generated by the cluster of $\alpha$-particles. As an example, I discuss the properties of the neighboring nuclei $^{12}$C and $^{13}$C, in particular energies, electromagnetic transitions and form factors.

\section{Algebraic Cluster Model}

The Algebraic Cluster Model (ACM) was introduced to describe the relative motion of cluster systems \cite{ACM,PS,PPNP}. It is based on the general criterion to introduce a 
$U(\nu+1)$ spectrum generating algebra for a bound-state problem with $\nu$ degrees 
of freedom. The relevant degrees of freedom of a system of $n$-body clusters, after removing the center-of-mass motion, are given by the $n-1$ relative Jacobi coordinates 
\ba
\vec{\rho}_{k} &=& \frac{1}{\sqrt{k(k+1)}}  
\left( \sum_{j=1}^{k} \vec{r}_{j} - k \vec{r}_{k+1} \right) ~,
\label{Jacobi}
\ea
with $k=1,\ldots,n-1$, and their conjugate momenta, $\vec{p}_{k}$. 
The building blocks of the ACM consist of a vector boson for each relative coordinate and momentum to which a scalar boson is added under the constraint that the total number of bosons is conserved.  
In the ACM, cluster states are described in terms of a system of $N$ interacting 
bosons with angular momentum and parity $L^P=1^-$ (dipole or vector bosons,  $b^{\dagger}_{k}$) and $L^P=0^+$ (monopole or scalar bosons, $s^\dagger$). The $3(n-1)$ components of the vector bosons together with the scalar boson span a $(3n-2)$-dimensional space with group structure $U(3n-2)$. The many-body states are classified according to the totally symmetric irreducible representation $[N]$ of $U(3n-2)$, where $N$ represents the total number of bosons $N=n_s+\sum_k n_k$. 

\subsection{Case study: $^{12}$C}

For the case of $^{12}$C as a cluster of three identical $\alpha$ particles the Hamiltonian has to be invariant with respect to the permutation symmetry $S_3$, {\it i.e.} the interchange of any of the $\alpha$ particles. The classical limit of such a  Hamiltonian has a triangular equilibrium configuration in which the relative distance of the three $\alpha$-particles to the center-of-mass is the same $r_i=\beta$, and the coordinates are given by $(\beta,\frac{\pi}{2},0)$, $(\beta,\frac{\pi}{2},\frac{2\pi}{3})$ and $(\beta,\frac{\pi}{2},\frac{4\pi}{3})$ \cite{ACM}. For the case of triangular symmetry the eigenstates of the ACM can be classified according the irreducible representations (irreps) of the point group ${\cal D}_{3h}$. Since ${\cal D}_{3h} \sim {\cal D}_3 \times P$, the transformation properties under ${\cal D}_{3h}$ can also be labeled by parity $P$ and the representations of ${\cal D}_3 \sim S_3$. 

The corresponding rotation-vibration excitations are characterized by the quantum numbers
\begin{equation} 
\mid (v_{1},v_{2}^{\ell});K,L_{t}^{P},M \rangle ~.
\end{equation}
Here $(v_1,v_2^\ell)$ denote the vibrational quantum numbers: $v_1$ is a one-dimensional stretching vibration and $v_2^\ell$ a two-dimensional vibration consisting of an asymmetric stretching and a bending vibration. The label $\ell$ is the proyection of the vibrational angular momentum on the symmetry axis \cite{BDL}. The label $K$ is the projection of the angular momentum $L$ on the symmetry axis, $M$ is the magnetic substate and $t$ denotes the permutation symmetry under $S_3$.  
The energy spectrum is given by 
\ba
E &=&\omega_{1}(v_{1}+\frac{1}{2}) + \omega_{2}(v_{2}+1) 
+ \kappa_{1} L(L+1) + \kappa_2 (K \mp 2\ell)^2 ~.
\ea
Here $\omega_1$ denotes the frequency of the symmetric stretching (or breathing) vibration, $\omega_2$ that of the two-dimensional vibration and $\kappa_1$ is proportional to the inverse of the moment of inertia. The last term proportional to $\kappa_2$ contains the effects of the Coriolis interaction \cite{ACM}.

\begin{figure}[t]
\begin{minipage}{.5\linewidth}
\includegraphics[width=\linewidth]{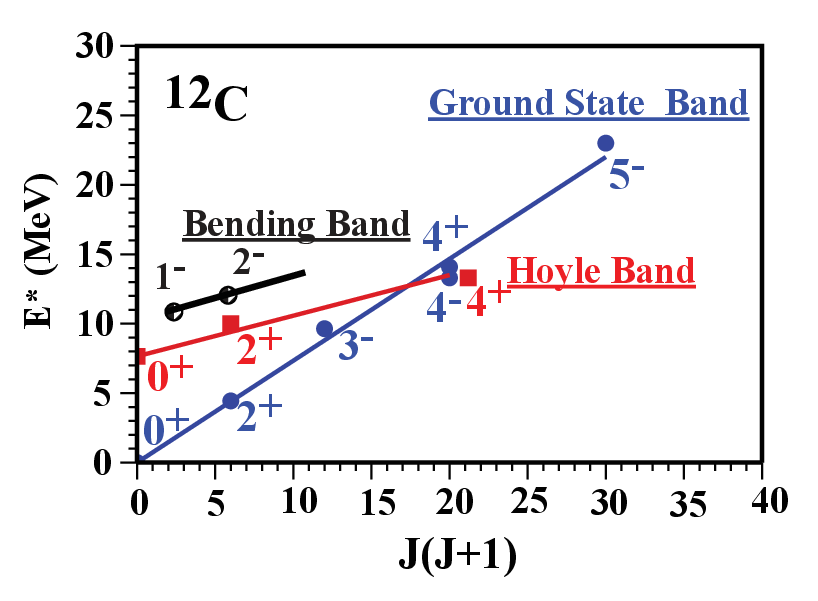}
\end{minipage}\hfill
\begin{minipage}{.5\linewidth}
\includegraphics[width=\linewidth]{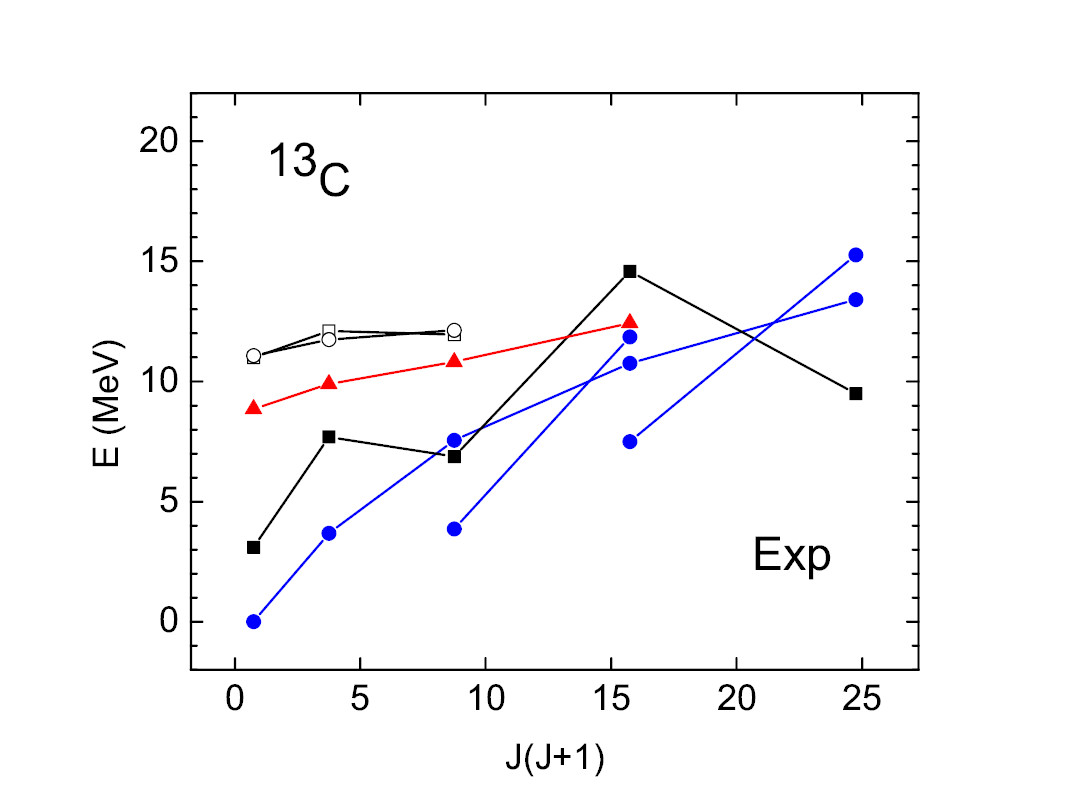}
\end{minipage}\hfill
\caption{Left: Rotational band structure in $^{12}$C \cite{Marin} (left) 
and $^{13}$C \cite{C13} (right).}
\label{RotBands}
\end{figure}

Fig.~\ref{RotBands} shows the band structure of the experimentally observed cluster states in $^{12}$C. The ground state band $(v_1,v_2^\ell)=(0,0^0)$ consists of positive parity states with $K=0$ and negative parity states with $K=3$ which merged into a single rotational sequence $L^P=0^{+}$, $2^{+}$, $3^{-}$, $4^{\pm}$, $5^{-}$, all of which have been identified experimentally (blue) \cite{Marin}. In the ACM, the Hoyle band with $(1,0^0)$ has the same rotational structure, of which only the states with positive parity, $0^+$, $2^+$ and  $4^+$, have been found so far (red). Nuclear effective field theory calculations suggested a bent-arm configuration for the Hoyle state \cite{Epelbaum} rather than a triangular configuration. The measurement of these "missing" negative parity states is crucial to determine the geometric configuration of the Hoyle state. 
Finally, in Fig.~\ref{RotBands} there are some candidates of the rotational band built on top of the bending vibration with $(0,1^1)$ (black).

\section{Cluster Shell Model (CSM)}

The cluster shell model was introduced \cite{CSM1,CSM2} to describe nuclei composed of $n$ $\alpha$-particles plus additional nucleons, denoted as $A=4n + x$ nuclei. The CSM combines cluster and single-particle degrees of freedom, and is very similar in spirit as the Nilsson model \cite{Nilsson}, but in the CSM the odd nucleon moves in the deformed field generated by the (collective) cluster degrees of freedom. 
The Hamiltonian is given by
\ba
H = T + V(\vec{r}) + V_{\rm so}(\vec{r}) + \frac{1}{2}(1+\tau_3) V_{\rm C}(\vec{r}) ~,
\label{hcsm}
\ea
{\it i.e.} the sum of the kinetic energy, the interaction between the $\alpha$-particle and the nucleon, a spin-orbit interaction and, for an odd proton, a Coulomb potential. The label $\tau_3=2t_3$ is twice the projection of isospin of the extra nucleon and has the value $+1$ for a proton and $-1$ for a neutron.

\begin{figure}[htb]
\centering
\begin{minipage}{.3\linewidth}
\centerline{\includegraphics[scale=0.3]{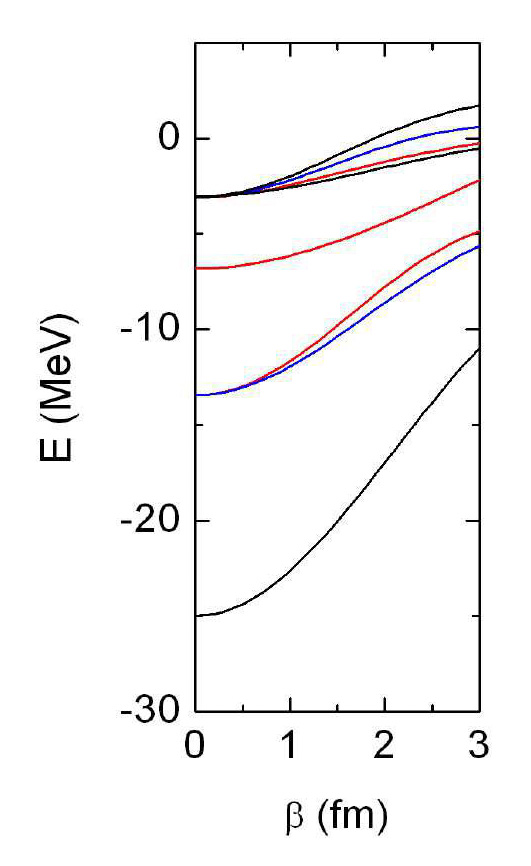}}
\end{minipage}\hfill
\begin{minipage}{.7\linewidth}
\centerline{\includegraphics[scale=0.55]{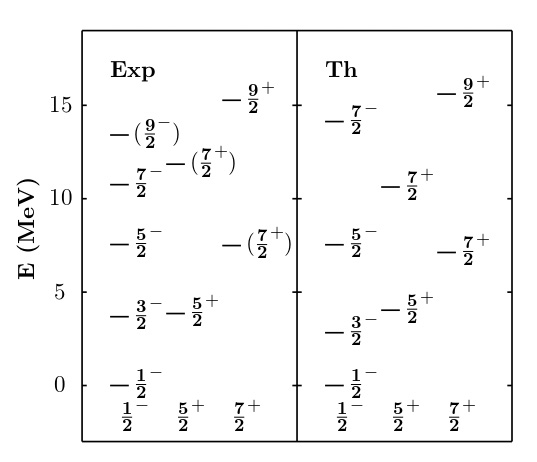}}
\end{minipage}
\caption[]{\label{csm3}Left panel: Single-particle energies in a cluster potential with ${\cal D}_{3h}^{\prime}$ symmetry. The single-particle levels are labeled by $E_{1/2}$ (black), $E_{5/2}$ (red) and $E_{3/2}$ (blue). Right panel: Comparison between experimental and theoretical energies for the ground state band assigned to the representation $E_{5/2}$. The values of $K^P$ are given at the bottom of the figure.}
\end{figure}

\subsection{Case study: $^{13}$C}

The $^{13}$C nucleus is described in the CSM as a cluster of three $\alpha$ particles and an extra neutron ($n=3$ and $x=1$). For the case of triangular symmetry the eigenstates of the CSM Hamiltonian of Eq.~(\ref{hcsm}) can be classified according to the doubly degenerate spinor representations of the double point group ${\cal D}'_{3h} $\cite{Herzberg}: $\Omega=E_{1/2}$, $E_{5/2}$ and $E_{3/2}$  or, in the notation of Ref.~\cite{C13}, as $\Omega=E_{1/2}^{(+)}$, $E_{1/2}^{(-)}$ and $E_{3/2}$, respectively. The Hamiltonian of the CSM is solved in the intrinsic, or body-fixed, system adopting a set of symmetry-adapted basis states \cite{EPJST}.

The rotational states are labeled by the angular momentum $J$, parity $P$ and its projection $K$ on the symmetry axis, $| J^P K \rangle$. The allowed values of $K^P$ for each one of the spinor representations are given by \cite{EPJST,PLB}
\ba
\begin{array}{lcl}
\Omega=E_{1/2} &:& K^P = \frac{1}{2}^+, \frac{5}{2}^-, \frac{7}{2}^-, \ldots \\ 
\Omega=E_{5/2} &:& K^P = \frac{1}{2}^-, \frac{5}{2}^+, \frac{7}{2}^+, \ldots \\ 
\Omega=E_{3/2} &:& K^P = \frac{3}{2}^+, \frac{3}{2}^-, \frac{9}{2}^+, \ldots 
\end{array}
\ea
with $J=K$, $K+1$, $\ldots$. The ground state band in $^{13}$C has $E_{5/2}$ symmetry under ${\cal D}'_{3h}$ and consists of several rotational bands labeled by $K^P=\frac{1}{2}^-$, $\frac{5}{2}^+$, $\frac{7}{2}^+$, $\ldots$ \cite{C13}. 

The rotational excitation energies are given by \cite{C13,EPJST}
\ba
E_{\rm rot} \;\approx\; \frac{1}{2{\cal I}} \left[ J(J+1) - 2K^2
+ \delta_{K,1/2} \, a_{\Omega} (-1)^{J+1/2} \left( J+\frac{1}{2} \right) \right] ~,
\label{erot}
\ea
where the last term denotes the Coriolis mixing. 

The left panel of Fig.\ref{csm3} shows the single-particle level splitting in the CSM as a function of $\beta$, the relative distance of the $\alpha$ particles with respect to the center of mass. The energetically most favorable level for the seventh neutron in $^{13}$C has $\Omega=E_{5/2}$ symmetry and contains rotational bands with $K^P=\frac{1}{2}^-$ and 
$\frac{5}{2}^+$, $\frac{7}{2}^+$ all of which have been identified (right pannel). 

The right panel of Fig.~\ref{RotBands} shows the rotational band structure in $^{13}$C which consists of rotational states with $\Omega=E_{5/2}$ (blue) and  $E_{1/2}$ 
symmetry (black). The $K^P=\frac{1}{2}^+$ band has a large decoupling parameter (black). The red states form a rotational band which was assigned as the analogue of the Hoyle band \cite{C13}. The moment of inertia of this band is very similar to that of the Hoyle band in $^{12}$C

\section{Coulomb form factors}

Another important test of the wave functions of $^{12}$C and $^{13}$C is provided by electromagnetic form factors and transitions. For a system consisting of three $\alpha$-particles in a triangular geometry and a single nucleon the charge distribution is taken be the sum of a Gaussian-like distribution for the three $\alpha$-particles and a point-like distribution for the extra nucleon
\ba
\rho_{\rm ch}(\vec{r}) &=& \rho_{\rm ch}^{\rm c}(\vec{r}) + \rho_{\rm ch}^{\rm sp}(\vec{r}) 
\nonumber\\
&=& \frac{Ze}{3} \left( \frac{\alpha }{\pi} \right)^{3/2} 
\sum_{i=1}^{3} \mbox{e}^{-\alpha( \vec{r}-\vec{r}_{i})^{2}} 
+ \tilde{e} \, \delta(\vec{r}-\vec{r}_{\rm sp}) ~.
\ea
Here $Ze$ is the electric charge of the $^{12}$C nucleus and $\tilde{e}$ denotes the effective charge of the extra nucleon. Since for $^{13}$C the extra nucleon is a neutron the single-particle effective charge is taken to be $\tilde{e}=0$. 
The Coulomb (or longitudinal) form factor is given by the matrix element of the Fourier transform of the charge distribution. The Coulomb form factor for vibrationally elastic excitations from the ground state is given by \cite{ACM,C13,PLB}
\ba
F_{CL}(q; 0^+,0 \rightarrow L^P,K) 
= \sqrt{4\pi} \, j_L(q\beta) \, \mbox{e}^{-q^2/4\alpha} \, Y_{LK}(\frac{\pi}{2},0) \, \sqrt{\frac{2}{1+\delta_{K0}}} \, \delta_{K,3\kappa} ~,
\ea
with $P=(-)^L$ for $^{12}$C, and by
\ba
F_{CL}(q;\frac{1}{2}^-,\frac{1}{2} \rightarrow J^{P'},\pm K + \frac{1}{2}) = \left< \frac{1}{2},\frac{1}{2},L,\pm K \left| J,\pm K +\frac{1}{2}\right>\right.  
F_{CL}(q; 0^+,0 \rightarrow L^P,\pm K) ~,
\ea
with $J=L\pm \frac{1}{2}$ and $P'=-(-)^L$ for $^{13}$C. 

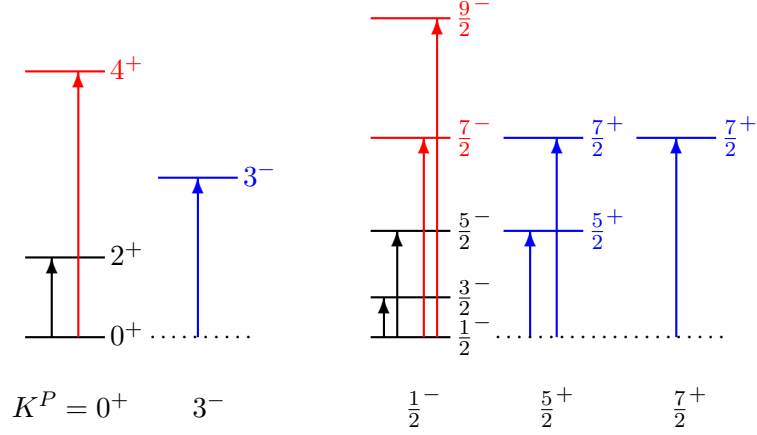
\begin{figure}
\centering
\setlength{\unitlength}{1pt}
\begin{picture}(350,200)(25,0)
\thicklines
\put( 50, 50) {\line(1,0){30}} 
\put( 50, 80) {\line(1,0){30}} 
\put( 50,150) {\color{red} \line(1,0){30}} 
\put( 60, 50) {\vector(0,1){30}} 
\put( 70, 50) {\color{red} \vector(0,1){100}} 
\put( 82, 47) {$0^+$}
\put( 82, 77) {$2^+$}
\put( 82,147) {\color{red} $4^+$}
\put(100,110) {\color{blue} \line(1,0){30}}
\put(115, 50) {\color{blue} \vector(0,1){60}} 
\put(132,107) {\color{blue} $3^-$}
\put( 45, 20) {$K^P=0^+$}
\put(113, 20) {$3^-$}
\multiput(98, 50)(4,0){10}{\circle*{1}}

\put(180, 50) {\line(1,0){30}} 
\put(180, 65) {\line(1,0){30}} 
\put(180, 90) {\line(1,0){30}} 
\put(180,125) {\color{red} \line(1,0){30}} 
\put(180,170) {\color{red} \line(1,0){30}} 
\put(185, 50) {\vector(0,1){15}} 
\put(190, 50) {\vector(0,1){40}} 
\put(200, 50) {\color{red} \vector(0,1){75}} 
\put(205, 50) {\color{red} \vector(0,1){120}} 
\put(212, 47) {$\frac{1}{2}^-$}
\put(212, 62) {$\frac{3}{2}^-$}
\put(212, 87) {$\frac{5}{2}^-$}
\put(212,122) {\color{red} $\frac{7}{2}^-$}
\put(212,167) {\color{red} $\frac{9}{2}^-$}

\put(230, 90) {\color{blue} \line(1,0){30}} 
\put(230,125) {\color{blue} \line(1,0){30}} 
\put(240, 50) {\color{blue} \vector(0,1){40}} 
\put(250, 50) {\color{blue} \vector(0,1){75}} 
\put(262, 87) {\color{blue} $\frac{5}{2}^+$}
\put(262,122) {\color{blue} $\frac{7}{2}^+$}
\put(280,125) {\color{blue} \line(1,0){30}} 
\put(295, 50) {\color{blue} \vector(0,1){75}} 
\put(312,122) {\color{blue} $\frac{7}{2}^+$}

\put(193, 20) {$\frac{1}{2}^-$}
\put(243, 20) {$\frac{5}{2}^+$}
\put(293, 20) {$\frac{7}{2}^+$}
\multiput(228, 50)(4,0){22}{\circle*{1}}
\end{picture}
\caption[]{Non-vanishing Coulomb form factors for $C2$ (black), $C3$ (red) and $C4$ (blue) vibrationally elastic excitations from the ground state of $^{12}$C (left)  and $^{13}$C (right).}
\label{Coulomb}
\end{figure}

As a consequence, ratios of these vibrationally diagonal longitudinal form factors are given by ratios of Clebsch-Gordan coefficients. These relations are similar, but not identical, to the well-known Alaga rules in the collective geometric model \cite{Alaga}. 
For a given multipole the $CL$ Coulomb form factors in $^{12}$C and $^{13}$C all have the same $q$ dependence, and the summed strength for the excitation from the ground state to excited states is the same in both nuclei
\ba
\left| F_{CL}(q;0^+,0 \rightarrow L^P,K) \right|^2_{^{12}C} &=& \sum_{JK'} \left| F_{CL}(q;\frac{1}{2}^-,\frac{1}{2} \rightarrow J^{P'},K')  \right|^2_{^{13}C} ~,
\label{ffC12C13}
\ea
with $J=L\pm \frac{1}{2}$ and $K'=\pm K +\frac{1}{2}$. The situation is illustrated in Fig.~\ref{Coulomb} for the multipoles with $C2$, $C3$, $C4$. In $^{12}$C only a single state is excited for multipoles $C2$, $C3$ and $C4$. Even though in $^{13}$C the strength is fragmented, it is only fragmented over a limited number of excited states, two states for $C2$ and $C4$ and 3 states for $C3$. 

\begin{figure}[!]
\centering
\includegraphics[scale=0.5]{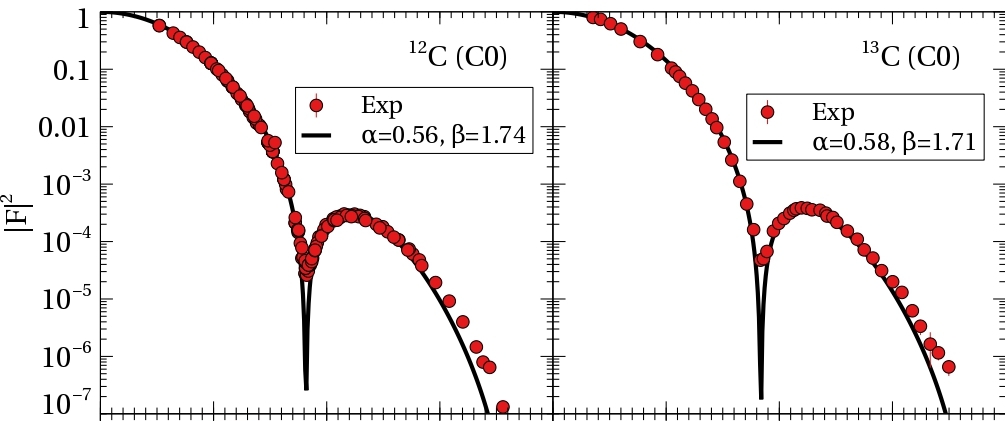}
\includegraphics[scale=0.5]{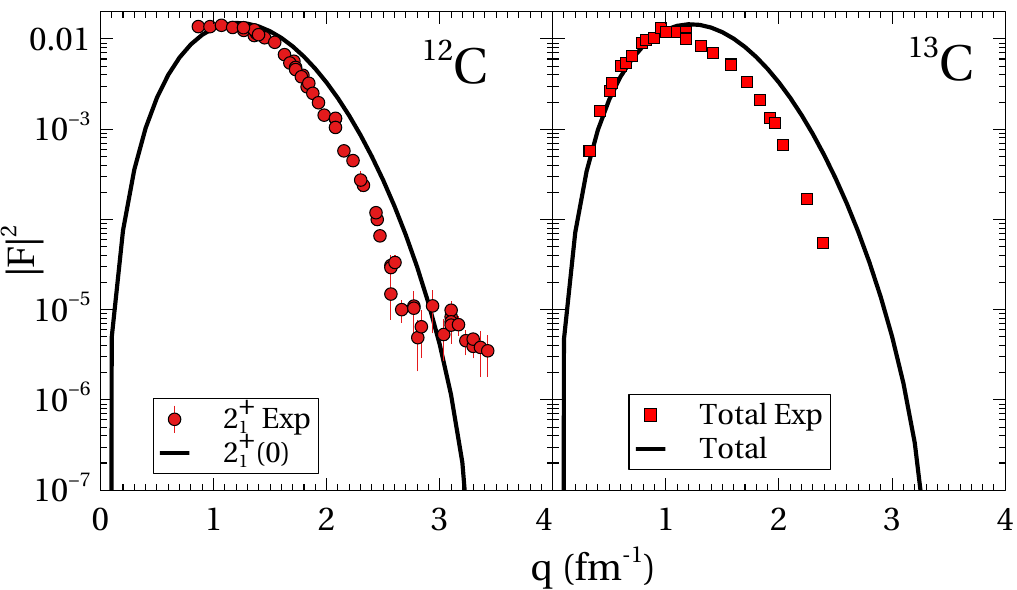}
\caption[]{Elastic C0 form factor in $^{12}$C (top left) and $^{13}$C(top right) and C2 form factor in $^{12}$C (bottom left) and $^{13}$C (bottom right). For $^{13}$C the squares represent the sum of the C2 form factors to the states $5/2_1^-$ and $3/2_1^-$. The experimental data are taken from \cite{crannell1,crannell2,Millener}. 
\label{CLff}}
\end{figure}

The Coulomb form factors depend on $\alpha$ and $\beta$. The value of $\alpha$ is determined from the radius of the $\alpha$ particle to be $\alpha=0.56$ 1/fm$^2$, and that of $\beta$ from the minimum in the elastic form factor  of $^{12}$C as $1.82$ fm. 
Fig.~\ref{CLff} shows a comparison of the elastic form factors in $^{12}$C and $^{13}$C as well as the summed $C2$ strength in both nuclei. In $^{12}$C only a single state is excited for the $C2$ multipole. In $^{13}$C the $C2$ strength is fragmented over two states, $J^P=5/2_1^-$ and $3/2_1^-$. The experimental data in Figs.~\ref{CLff} confirm the similarity between the $C0$ and $C2$ form factors in $^{12}$C and $^{13}$C as a consequence of the cluster structure in both nuclei. The overall agreement with experimental data is good. At higher $q$ values the form factors become more sensitive to short-range nuclear properties such as meson-exchange currents \cite{Hicks} which are not taken into account in the present analysis. 

\begin{figure}[!]
\centering
\includegraphics[scale=0.5]{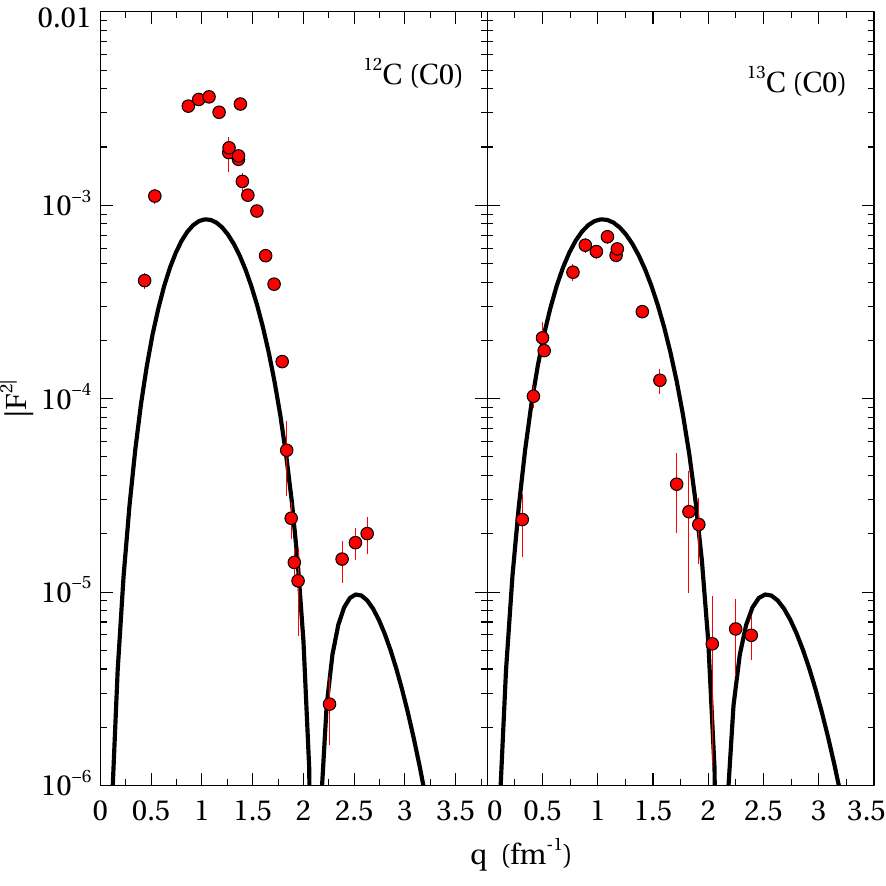}
\caption[]{Inelastic C0 form factor to the $0_2^+$ Hoyle state in $^{12}$C (left) and the $1/2_2^-$ state at 8.86 MeV in $^{13}$C (right). The experimental data are taken from \cite{crannell1,crannell2,Millener,strehl}. The solid line shows the calculated form factor as taken from \cite{ACM}.}
\label{ffhoyle}
\end{figure}

Similarly, the $C0$ form factors of the Hoyle state and its analogue in $^{13}$C are expected to be the same \cite{PLB}. Figure~\ref{ffhoyle} shows the form factors of the Hoyle state in $^{12}$C and the $J^P=1/2^-$ state at $8.860$ MeV  in $^{13}$C. The $q$ dependence is consistent with experimental data which suggests to interpret the $J^P=1/2^-$ state as the analogue of the Hoyle state in agreement with the previous analysis based on energy systematics as shown in Fig.~\ref{RotBands}, even though the strength of the Hoyle is underpredicted. 

\section{Electric  transitions}

The transition probabilities correspond to the long wavelength limit of the Coulomb form factors. Therefore, the $B(EL)$ values for excitations from the ground state of $^{12}$C and $^{13}$C satisfy the same type of correlations as the Coulomb form factors in Eq.~(\ref{ffC12C13}). Unlike the form factors, the $B(EL)$ values and quadrupole moments only depend on $\beta$ and are given by
\ba
B(EL;0^+,0 \rightarrow L^P,K) &=& \frac{(Ze)^2}{4\pi} c_{LK}^2 \beta^{2L} ~,
\nonumber\\
c_{LK}^2 &=& \delta_{K,3\kappa} \, \frac{8\pi}{1+\delta_{K0}} 
\left[ Y_{LK}(\frac{\pi}{2},0) \right]^2  ~,
\nonumber\\
Q_{2^+,0} &=& \frac{2}{7} Ze  \beta^2 ~,
\label{belc12}
\ea
for $^{12}$C with $c_{20}^2 = 5/4$, $c_{33}^2 = 35/8$ and $c_{40}^2 = 81/64$,  
and by
\begin{eqnarray}
B(EL;(\frac{1}{2}^-,\frac{1}{2} \rightarrow J^{P'},\pm K + \frac{1}{2})  &=& 
\left|\left< \frac{1}{2},\frac{1}{2},L,\pm K \left| J,\pm K +\frac{1}{2}\right>\right|^2 \right.  B(EL;0^+,0 \rightarrow L^P,K)  ~,
\nonumber\\
Q_{5/2^-,1/2} \;=\; \frac{10}{7} \, Q_{3/2^-,1/2} &=& Q_{2^+,0} ~,
\label{belc13}
\end{eqnarray}
for $^{13}$C. 
Table~\ref{BELs} shows that to a good approximation the experimental data in $^{12}$C and $^{13}$C can be understood by the correlations shown in Eqs.~(\ref{belc12}) and (\ref{belc13}). 

\begin{table}
\centering
\caption[]{Electric transition probabilities and quadrupole moments in $^{12}$C and $^{12}$C . \label{BELs}}
\vspace{10pt}
\begin{tabular}{lcccl}
\hline
\noalign{\smallskip}
&& ACM/CSM & Exp & \\
\noalign{\smallskip}
\hline
\noalign{\smallskip}
$^{12}$C & $B(E2;2^+,0 \rightarrow 0^+,0)$ & $7.8$ & $7.63 \pm 0.19$ & e$^2$fm$^4$ \\
& $B(E3;3^-,3 \rightarrow 0^+,0)$ & $65$ & $104 \pm 14$ & e$^2$fm$^6$ \\
& $B(E4;4^+,0 \rightarrow 0^+,0)$ & $48$ & & e$^2$fm$^8$ \\
& $Q_{2_1^+,0}$ & $+5.7$ & $+5.97 \pm 0.30$ & e fm$^2$ \\
\noalign{\smallskip}
\hline
\noalign{\smallskip}
$^{13}$C & $B(E2;\frac{3}{2}^-,\frac{1}{2} \rightarrow \frac{1}{2}^-,\frac{1}{2})$ & $7.8$ & $6.4 \pm 1.5$ & e$^2$fm$^4$ \\
& $B(E2;\frac{5}{2}^-,\frac{1}{2} \rightarrow \frac{1}{2}^-,\frac{1}{2})$ & $7.8$ & $5.6 \pm 0.4$ & e$^2$fm$^4$ \\
& $B(E3;\frac{5}{2}^+,\frac{5}{2} \rightarrow \frac{1}{2}^-,\frac{1}{2})$ & $65$ & $100 \pm 40$ & e$^2$fm$^6$ \\
& $Q_{5/2^-,1/2}$ & $+5.7$ & & e fm$^2$ \\
& $Q_{3/2^-,1/2}$ & $+4.0$ & & e fm$^2$ \\
\noalign{\smallskip}
\hline
\end{tabular}
\end{table}

\section{Summary and Conclusions}

In this contribution I presented a review of two algebraic approaches to clustering phenomena in light nuclei: the ACM for $\alpha$-conjugate nuclei with $A=4n$ and the CSM for nuclei with $A=4n+x$. The methods were illustrated with an application to $^{12}$C and $^{13}$C. Particular emphasis was made on the correlations between electromagnetic properties in both nuclei. It was shown that Coulomb form factors and transition probabilities are related by Clebsch-Gordan coefficients. A comparison with experimental data showed that the available data can be explained very well by these correlations which are a consequence of the underlying cluster structure in these nuclei. In addition, based on an analysis of energy systematics and Coulomb $C0$ form factors it was suggested to identify the $J^P=1/2^-$ state at 8.86 MeV in $^{13}$C as the analogue of the Hoyle state. 

The simultaneous analysis of energy spectra and electromagnetic properties in $^{12}$C and $^{13}$C shows strong evidence for the underlying cluster structure as three 
$\alpha$-particles in a triangular configuration plus an additional neutron which moves in the deformed field generated by the cluster. Currently, we are studying $\alpha$-cluster nuclei with $x=2$ extra nucleons to investigate to what extent the cluster structure of $\alpha$-particles persists under the addition of two nucleons. As a first example, we are studying the isospin triplet $^{10}$Be -- $^{10}$B -- $^{10}$C as a cluster of $n=2$ $\alpha$-particles and $x=2$ nucleons. Some preliminary results for $^{10}$Be were presented in Ref.~\cite{be10}. 

\section*{Acknowledgments}

This work was supported in part by research grant IN115426 from PAPIIT-DGAPA, UNAM, Mexico.

\end{document}